\documentclass[%
 aip,
 amsmath,amssymb,
 reprint,%
]{revtex4-1}

\usepackage{graphicx}
\usepackage{dcolumn}
\usepackage{bm}

\usepackage[utf8]{inputenc}
\usepackage[T1]{fontenc}
\usepackage{mathptmx}
\usepackage{etoolbox}

\makeatletter
\def\@email#1#2{%
 \endgroup
 \patchcmd{\titleblock@produce}
  {\frontmatter@RRAPformat}
  {\frontmatter@RRAPformat{\produce@RRAP{*#1\href{mailto:#2}{#2}}}\frontmatter@RRAPformat}
  {}{}
}%
\makeatother
\begin{document}

\title{Impact of noise on the instability of spiral waves in stochastic 2D mathematical models of human atrial fibrillation}
\author{Euijun Song}
 \email{drjunsong@gmail.com}
\affiliation{ 
Yonsei University College of Medicine, Seoul, Republic of Korea.\\
Present: Independent Researcher, Gyeonggi, Republic of Korea.
}

\date{\today}

\begin{abstract}
Sustained spiral waves, also known as rotors, are pivotal mechanisms in persistent atrial fibrillation (AF). Stochasticity is inevitable in nonlinear biological systems such as the heart; however, it is unclear how noise affects the instability of spiral waves in human AF. This study presents a stochastic two-dimensional mathematical model of human AF and explores how Gaussian white noise affects the instability of spiral waves. In homogeneous tissue models, Gaussian white noise may lead to spiral-wave meandering and wavefront break-up. As the noise intensity increases, the spatial dispersion of phase singularity (PS) points increases. This finding indicates the potential AF-protective effects of cardiac system stochasticity by destabilizing the rotors. By contrast, Gaussian white noise is unlikely to affect the spiral-wave instability in the presence of localized scar or fibrosis regions. The PS points are located at the boundary or inside the scar/fibrosis regions. Localized scar or fibrosis may play a pivotal role in stabilizing spiral waves regardless of the presence of noise. This study suggests that fibrosis and scars are essential for stabilizing the rotors in stochastic mathematical models of AF. Further patient-derived realistic modeling studies are required to confirm the role of scar/fibrosis in AF pathophysiology.
\end{abstract}

\maketitle

\section{Introduction}\label{sec1}

Atrial fibrillation (AF) is the most common cardiac arrhythmia characterized by chaotic electrical wave propagation, and is associated with mortality and morbidity \cite{Iwasaki:2011}. AF causes electrical and structural remodeling of the atrial tissues, evolving from paroxysmal AF to persistent AF (PeAF). Although the mechanisms of PeAF are poorly understood, recent studies suggest that PeAF is driven by sustained spiral waves (“rotors” or “reentrant drivers”) localized within spatially compacted regions \cite{Narayan:2012}. The core of a spiral wave, known as a spiral-wave tip, can be mathematically described as a phase singularity (PS) or a topological defect \cite{Gray:1998}, where the depolarizing wavefront and the repolarizing wavetail intersect \cite{Jalife:2000}. Cardiac computational modeling approaches have been widely used to study the complex spiral wave dynamics of human AF. In homogeneous atrial tissue models, the electrical remodeling conditions of PeAF can sustain stable rotor meandering in spatially compacted regions \cite{Pandit:2005}. In the presence of electrophysiological heterogeneities, rotors are frequently found in fibrotic regions or at the boundaries between fibrotic and non-fibrotic tissues \cite{Deng:2017,Roney:2016,Zahid:2016}. Additionally, it has been known that pinned spiral waves form extremely stable patterns with respect to external pacing and heterogeneities \cite{Bittihn:2010,Majumder:2020}. However, most computational models of human AF numerically solve deterministic partial differential equations (PDEs), ignoring the stochastic nature of complex biological systems.

Stochasticity is inevitable in complex biological systems such as gene regulatory networks, neuronal networks, and cardiac systems. It plays an important role in the dynamic behavior of nonlinear systems \cite{Buric:2007,Gammaitoni:1995}. For example, noise-induced stochastic resonance can be found in the FitzHugh–Nagumo model \cite{Pikovsky:1997}. In two-dimensional (2D) neuronal networks, certain thresholds of noise intensity can affect the formation and instability of spiral waves \cite{Ma:2010,Yao:2017}. However, it is unknown how noise affects the spiral wave dynamics in human AF models. Does noise affect spiral-wave instability in terms of the meandering of spiral-wave tips and wave break-up? Do PS points also localize near fibrotic regions in stochastic AF models? We do not know the answers to these questions based on experimental and clinical studies. Stochastic mathematical modeling of AF is essential for studying the effect of noise on the spiral wave dynamics in human AF.

\begin{figure}
\centering
\includegraphics[width=\columnwidth]{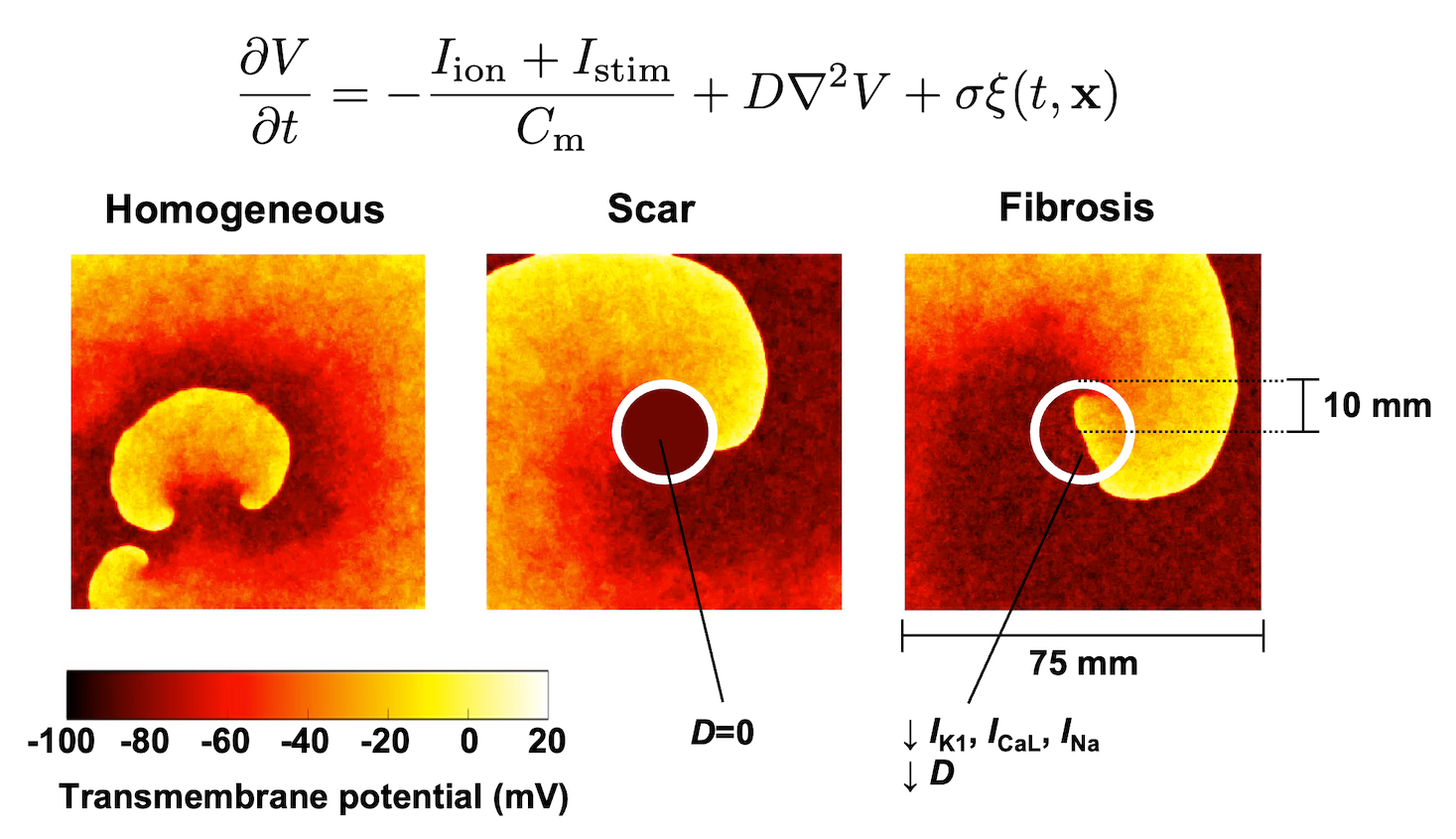}
\caption{Stochastic two-dimensional computational modeling of human atrial fibrillation. Spiral waves were numerically simulated on atrial tissues with and without localized scar and fibrosis regions (see Methods for details).}\label{Fig1}
\end{figure}

This study presents a stochastic 2D mathematical model of human AF by adding Gaussian white noise to the conventional deterministic reaction-diffusion equation. This study uses the Courtemanche-Ramirez-Nattel human atrial cell model \cite{Courtemanche:1998}; this atrial cell model is widely utilized in 2D and three-dimensional (3D) computational modeling of AF to study the spiral wave dynamics and personalize treatment strategies \cite{Azzolin:2023,Boyle:2019}. Using the stochastic mathematical model of AF, spiral waves are numerically simulated on 2D isotropic, homogeneous atrial tissues by varying noise intensity levels. To examine whether the PS points localize at fibrotic regions in the stochastic AF model, the spiral wave dynamics are further explored in the presence of localized scar or fibrosis regions (Figure \ref{Fig1}). This study shows that Gaussian white noise can lead to spiral-wave meandering and wavefront break-up in homogeneous atrial tissues, whereas localized scar or fibrotic regions can stabilize spiral waves without generating wavefront break-up.

\section{Methods}\label{sec2}

\subsection{Stochastic 2D mathematical modeling of AF}\label{subsec21}

In this section, I present a stochastic 2D mathematical model of human AF using the Courtemanche-Ramirez-Nattel human atrial cell model \cite{Courtemanche:1998}. The conventional deterministic 2D AF model can be described by the following reaction-diffusion equation \cite{Pandit:2005,Xie:2002}:
\begin{equation}
\frac{\partial V}{\partial t}=-\frac{I_{ion}+I_{stim}}{C_m}+D\nabla^{2}V
\end{equation}
where $V\left(t,\mathbf{x}\right)$ (mV) is the transmembrane potential, $I_{ion}$ (pA) is the total ionic current, $I_{stim}$ (pA) is the stimulus current, $C_{m}=100$ pF is the membrane capacitance, and $D$ (mm$^2$/ms) is the diffusion coefficient. The total ionic current is given by
\begin{multline}
I_{ion}=I_{Na}+I_{CaL}+I_{Kr}+I_{Ks}+I_{to}+I_{K1}+I_{Kur}+I_{NaK}\\+I_{NCX}+I_{pCa}+I_{Na,b}+I_{Ca,b}.
\end{multline}
The biophysical details of each ionic current can be found in Courtemanche et al \cite{Courtemanche:1998}. Here, I add a Gaussian white noise term to obtain the following stochastic PDE \cite{Jung:1995,Shardlow:2005}:
\begin{equation}
\label{eqn:spde}
\frac{\partial V}{\partial t}=-\frac{I_{ion}+I_{stim}}{C_m}+D\nabla^{2}V+\sigma\xi\left(t,\mathbf{x}\right)
\end{equation}
where $\sigma$ (mV) is the noise intensity and $\xi\left(t,\mathbf{x}\right)$ is the white noise satisfying
\begin{equation}
\left\langle\xi\left(t,\mathbf{x}\right)\right\rangle_{t}=0, \quad \left\langle\xi\left(t,\mathbf{x}\right), \xi\left(t',\mathbf{x}\right)\right\rangle_{t}=\delta\left(t-t'\right)
\end{equation}
for each $\mathbf{x}\in\mathbb{R}^2$. The above stochastic PDE (Eq. \ref{eqn:spde}) can be rewritten in the differential form as follows:
\begin{equation}
\label{eqn:spdediff}
dV=\left[-\frac{I_{ion}+I_{stim}}{C_m}+D\mathrm{\nabla}^{2}V\right]dt+\sigma dW\left(t,\mathbf{x}\right)
\end{equation}
where $W\left(t,\mathbf{x}\right)$ is the Wiener process satisfying $dW\left(t,\mathbf{x}\right)=\xi\left(t,\mathbf{x}\right)dt$, which is the differential form of the Brownian motion.

In the above stochastic PDE (Eq. \ref{eqn:spdediff}), the Gaussian white noise term is only added to membrane potential. However, this noise term can be viewed as a sum of the Gaussian noise of ion channels and membrane potential. Let's consider independent noise terms added to ion channels and membrane potential as follows:
\begin{eqnarray}
\label{eqn:spdediffdetail}
dV &=& -\frac{\sum_{i}{\left(I_{i} dt+\sigma_{i} dW_{i}\right)}}{C_m}+D\mathrm{\nabla}^{2}V dt+\sigma_{m} dW_{m} \nonumber \\
&=& -\frac{\sum_{i}{I_{i}}}{C_m} dt+D\mathrm{\nabla}^{2}V dt+\left(\sigma_{m} dW_{m}-\sum_{i}{\frac{\sigma_{i}}{C_{m}} dW_{i}}\right)
\end{eqnarray}
where $\sigma_{i}$ is the noise intensity of ion channel $I_{i}$, $W_{i}\left(t,\mathbf{x}\right)$ is the Wiener process related to the noise of ion channel $I_{i}$, $\sigma_{m}$ is the noise intensity of membrane potential, and $W_{m}\left(t,\mathbf{x}\right)$ is the Wiener process related to the membrane potential noise. Let $\sigma_{tot}$ be a total noise intensity defined as follows:
\begin{equation}
\sigma_{tot}=\sqrt{\sigma_{m}^{2} + \sum_{i}{\frac{\sigma_{i}^{2}}{C_{m}^{2}}}}.
\end{equation}
If noise terms are independent, the following sum of the Wiener processes is also the Wiener process:
\begin{equation}
W_{tot}=\frac{\sigma_{m}}{\sigma_{tot}} W_{m}+\sum_{i}{\frac{-\sigma_{i}/C_{m}}{\sigma_{tot}} W_{i}}.
\end{equation}
Therefore, the stochastic PDE (Eq. \ref{eqn:spdediffdetail}) can be rewritten as follows:
\begin{equation}
dV= -\frac{\sum_{i}{I_{i}}}{C_m} dt+D\mathrm{\nabla}^{2}V dt+\sigma_{tot} dW_{tot}.
\end{equation}
This is identical to the stochastic PDE presented before (Eq. \ref{eqn:spdediff}), indicating that this noise term can be viewed as a sum of the Gaussian noise of ion channels and membrane potential. Although the noise of ion channels and membrane potential may be neither Gaussian nor independent in the real situation, this study uses this simplified stochastic PDE (Eq. \ref{eqn:spdediff}) to examine the effects of noise on the spiral wave dynamics.

\subsection{Numerical simulation}\label{subsec22}

This study numerically solves the stochastic PDE (Eq. \ref{eqn:spdediff}) on a 2D isotropic, homogeneous domain of area 75×75 mm$^2$, consisting of a 2D lattice network of 300×300 atrial cells. The forward Euler method was used with a fixed time step of $\Delta t=0.02$ ms and a space step of $\Delta x=\Delta y=0.25$ mm. The Neumann (no-flux) boundary conditions were applied. The Laplacian $\nabla^{2}V$ was approximated using the five-point stencil. The increment of the Wiener process was numerically implemented in the Itô sense as
\begin{equation}
W\left(t+\Delta t,\mathbf{x}\right)-W\left(t,\mathbf{x}\right)\approx\sqrt{\Delta t}\ \eta
\end{equation}
where $\eta\sim\mathcal{N}\left(0,\ 1\right)$ is a Gaussian random number with a mean value of 0 and a standard deviation of 1 \cite{Guo:2012,Yao:2017}.

To reflect the electrical remodeling of PeAF, I reduced the L-type Ca$^{2+}$ current ($I_{CaL}$) by 70\%, the transient outward K$^{+}$ current ($I_{to}$) by 50\%, and the ultrarapid delayed rectifier K$^{+}$ current ($I_{Kur}$) by 50\%, and increased the inward rectifier K$^{+}$ current ($I_{K1}$) by 100\%, as described by Pandit et al \cite{Pandit:2005}. Two diffusion coefficients were tested: $D$=0.1 and 0.05 mm$^2$/ms, which produce conduction velocities of 0.43 and 0.27 m/s, respectively. $D$=0.1 mm$^2$/ms is a commonly used diffusion coefficient in 2D AF models, producing a physiological conduction velocity \cite{Pandit:2005,Xie:2002}. The noise intensity levels of $\sigma=$0, 1, 5, and 10 mV were tested. Spiral waves were initiated by applying the standard S1-S2 cross-field protocol, and the AF wave dynamics were studied for 5 s. The numerical simulation was performed using C++ code with OpenMP parallelization. The atrial cell model is publicly available at the CellML Physiome Project (https://models.physiomeproject.org).

\subsection{Modeling scar and fibrosis regions}\label{subsec23}

In addition to the 2D homogeneous model described in the previous section, the AF wave dynamics were simulated on inhomogeneous models in the presence of a localized scar or fibrotic regions. The scar and fibrotic regions were applied to the center of the 2D cardiac tissue with a radius of 10 mm (Figure \ref{Fig1}).
\begin{itemize}
\item \textit{Scar:} The scar was modeled as a nonconductive region ($D=0$). The no-flux boundary conditions were applied around the scar region.
\item \textit{Fibrosis:} In the fibrotic regions, I reduced $I_{K1}$ by 50\%, $I_{CaL}$ by 50\%, and the sodium current ($I_{Na}$) by 40\% to reflect the TGF-$\beta$1 fibrogenic signaling effects. The diffusion coefficient $D$ was also decreased by 30\% to represent gap junction remodeling \cite{Roney:2016,Zahid:2016}.
\end{itemize}

\begin{figure*}
\centering
\includegraphics[width=\textwidth]{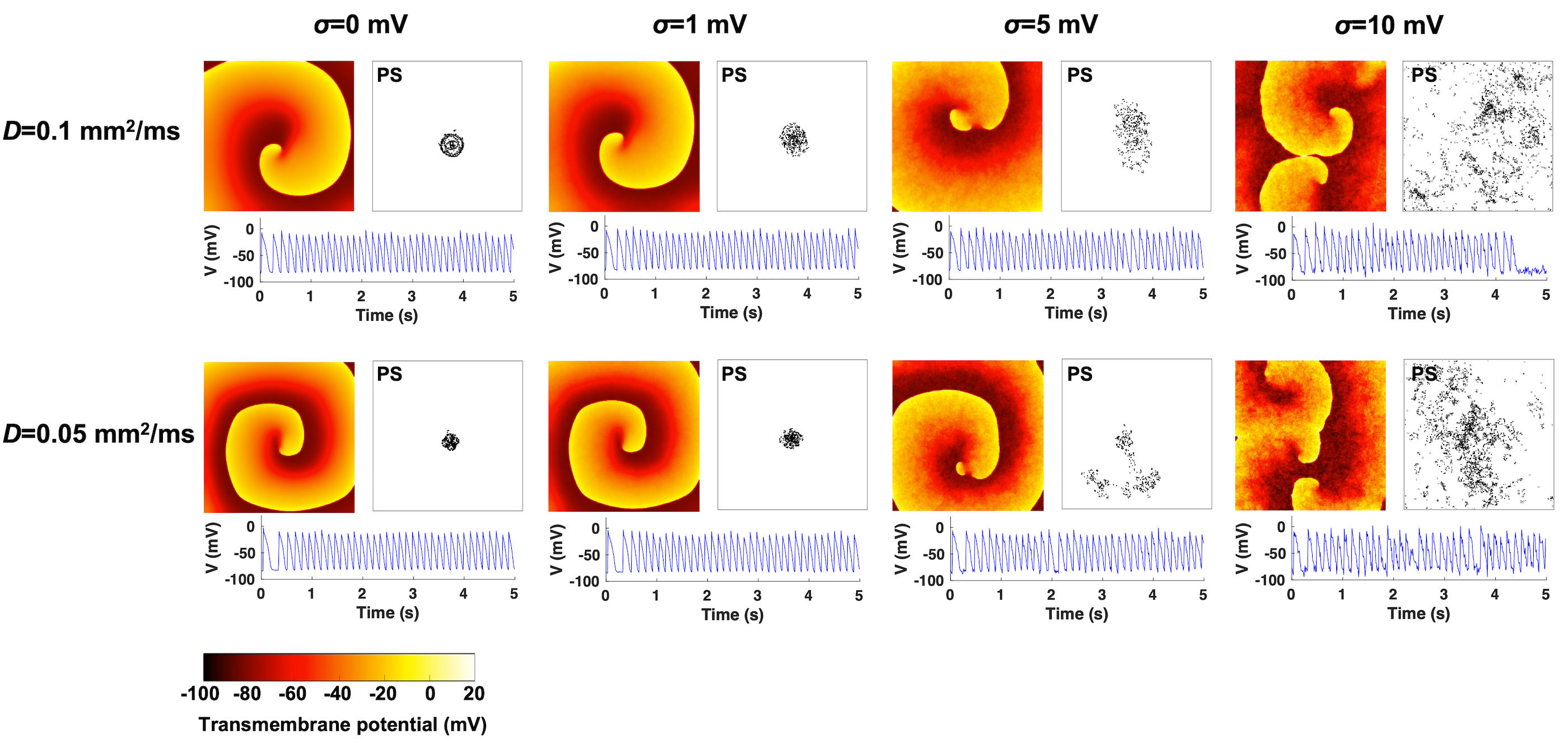}
\caption{Transmembrane potential maps and phase singularity (PS) plots for stochastic 2D atrial fibrillation simulations on homogeneous tissues. The simulations were performed for diffusion coefficients of $D$=0.1 and 0.05 mm$^2$/ms, and noise intensity levels of $\sigma$=0, 1, 5, and 10 mV. The PS points were computed during the whole fibrillation state, and the action potential signals were acquired at the node (50, 50).}\label{Fig2}
\end{figure*}

\begin{figure}
\centering
\includegraphics[width=\columnwidth]{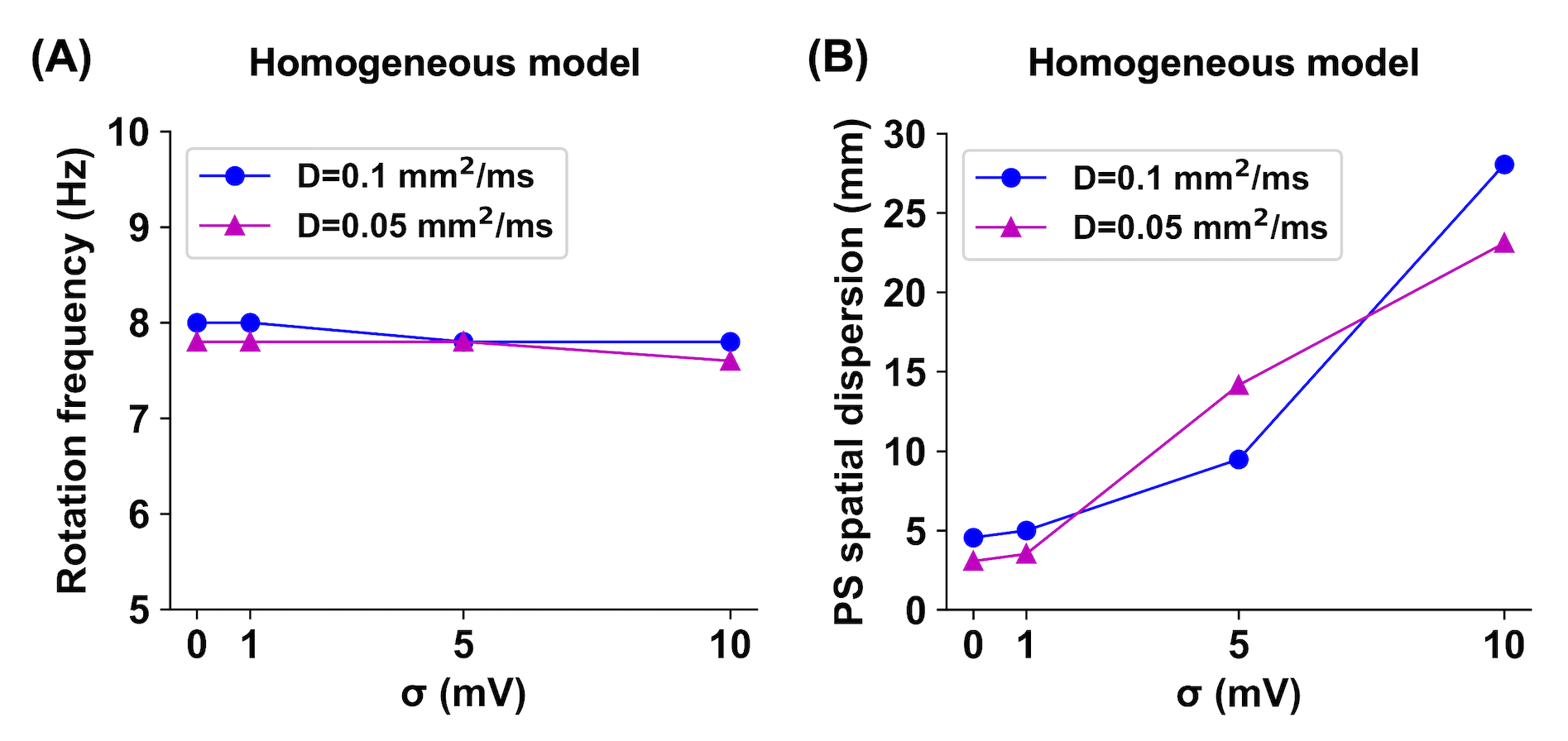}
\caption{Spiral wave rotation frequency (A) and phase singularity (PS) spatial dispersion (B) values for stochastic 2D atrial fibrillation simulations on homogeneous tissues, depending on the diffusion coefficients of $D$=0.1 and 0.05 mm$^2$/ms and the noise intensity levels of $\sigma$=0, 1, 5, and 10 mV.}\label{Fig3}
\end{figure}

\subsection{Analysis of spiral waves}\label{subsec24}

To analyze the spatiotemporal patterns of spiral waves, 2D maps of the transmembrane potential were generated with a sampling time step of 10 ms. PS points were identified using the method proposed by Iyer and Gray \cite{Iyer:2001}. The phase $\theta\left(t,\mathbf{x}\right)\in\left[-\pi,\pi\right]$ at each point is calculated as follows:
\begin{equation}
\theta\left(t,\mathbf{x}\right)=\arctan{\left[V\left(t+\tau,\mathbf{x}\right)-V_{mean}\left(\mathbf{x}\right),\ V\left(t,\mathbf{x}\right)-V_{mean}\left(\mathbf{x}\right)\right]}
\end{equation}
where $\tau=30$ ms is the time delay constant and $V_{mean}$ is the mean of the action potential for the whole fibrillation state. The PS points are identified if the winding number of phase field around the particular point is non-zero \cite{Bray:2001,Iyer:2001}:
\begin{equation}
\frac{1}{2\pi}\oint{\nabla\theta\cdot d\mathbf{r}}=\pm 1.
\end{equation}
To evaluate the spatial distribution of the PS points, the PS spatial dispersion was computed as the standard deviation of the PS points as follows:
\begin{equation}
\text{PS spatial dispersion}=\sqrt{\frac{\sum_{i}\left\|{PS}_{i}-{PS}_{mean}\right\|^{2}}{N_{PS}}}
\end{equation}
where ${PS}_{i}$ is the $i$th PS point, ${PS}_{mean}$ is the average location of the PS points, $N_{PS}$ is the number of PS points, and $\left\|\cdot\right\|$ is the $L^2$ norm. The PS spatial dispersion vanishes if the spiral-wave tip is consistent over time. The spiral wave rotation frequency is estimated as the maximum dominant frequency of the action potential signals acquired at the node (50, 50). All signal analyses were performed using MATLAB 2021b (MathWorks, Inc.).

\section{Results}\label{sec3}

\subsection{Noise-induced instability of spiral waves in homogeneous models}\label{subsec31}

First, human AF was numerically simulated on 2D homogeneous tissue models. Figure \ref{Fig2} shows the transmembrane potential maps and PS plots. With noise intensity levels of $\sigma$=0 and 1 mV, spiral waves were localized near the center of the atrial tissue, indicating sustained stable rotor dynamics. The maximum distance between the PS points was $<$20 mm. When a noise intensity level of $\sigma$=5 mV was applied, spiral waves continuously meandered ($>$30 mm) and wavefront break-up occurred. At a noise intensity level of $\sigma$=10 mV, spiral waves were largely meandered ($>$60 mm), and wavefront break-up also occurred. At a noise intensity level of $\sigma$=10 mV and a diffusion coefficient of $D$=0.1 mm$^2$/ms, the spiral waves were spontaneously terminated at 4.5 s. All the other cases showed sustained fibrillation states for $>$5 s. Sequences of the transmembrane potential maps are shown in Supplementary Figure S1.

Additionally, this study quantitatively determined how noise changes the spiral wave rotation frequency and PS spatial dispersion, as shown in Figure \ref{Fig3}. Noise changed the spiral wave frequency by only approximately $<$2.6\% (Figure \ref{Fig3}A). As the noise intensity level increased from 0 to 10 mV, the spiral wave rotation frequencies were 8.0, 8.0, 7.8, and 7.8 Hz for the $D$=0.1 mm$^2$/ms cases, and 7.8, 7.8, 7.8, and 7.6 Hz for the $D$=0.05 mm$^2$/ms cases, respectively. However, noise dramatically increased the PS spatial dispersion (Figure \ref{Fig3}B). As the noise intensity level increased from 0 to 10 mV, the PS spatial dispersions were 4.5, 5.0, 9.5, and 28.1 mm for the $D$=0.1 mm$^2$/ms cases, and 3.1, 3.5, 14.1, and 23.1 mm for the $D$=0.05 mm$^2$/ms cases, respectively. This result is consistent with the observations of noise-induced spiral-wave meandering and wavefront breakup, as shown in Figure \ref{Fig2}.

\begin{figure*}
\centering
\includegraphics[width=\textwidth]{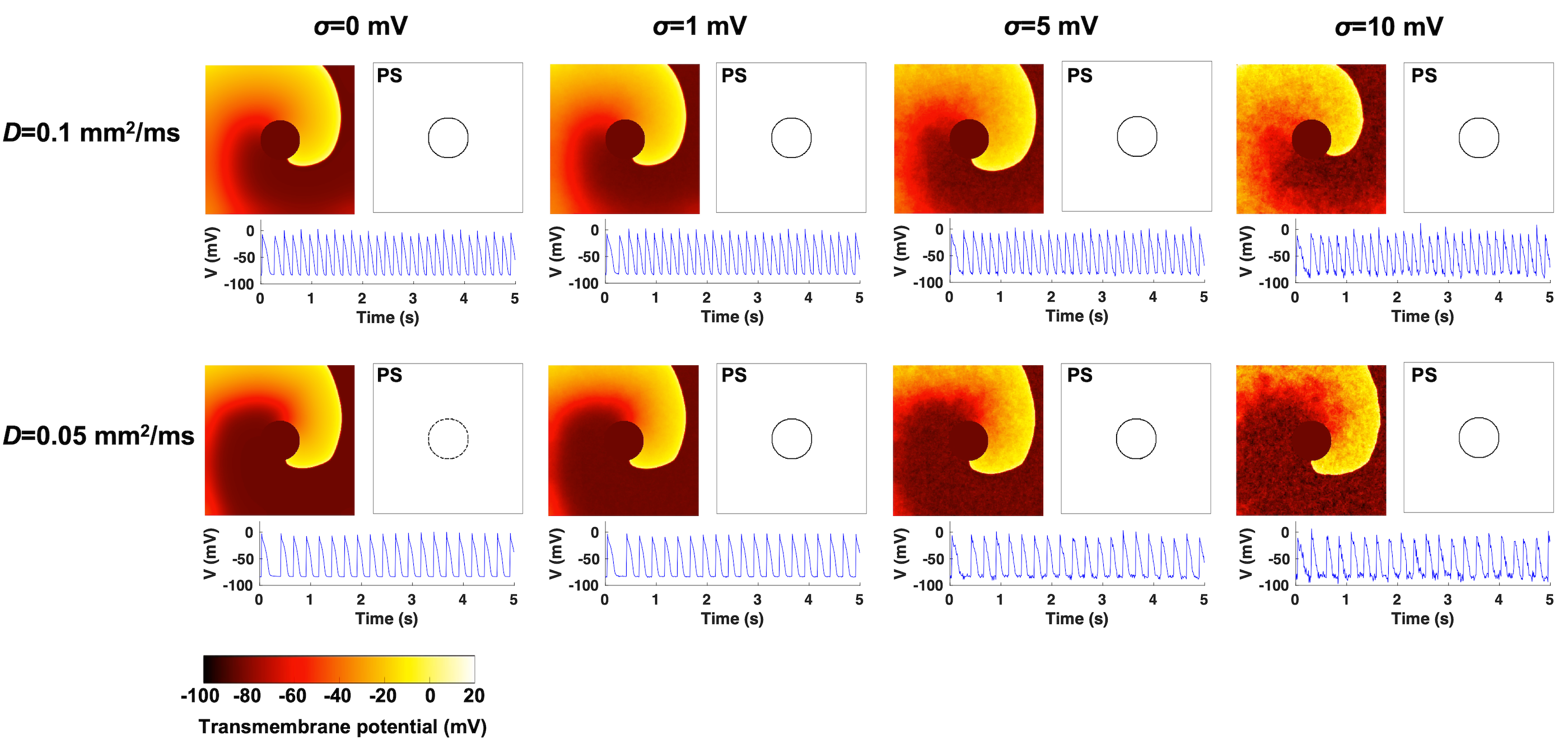}
\caption{Transmembrane potential maps and phase singularity (PS) plots for stochastic 2D atrial fibrillation simulations in the presence of scar regions. The simulations were performed for diffusion coefficients of $D$=0.1 and 0.05 mm$^2$/ms, and noise intensity levels of $\sigma$=0, 1, 5, and 10 mV. The PS points were computed during the whole fibrillation state, and the action potential signals were acquired at the node (50, 50).}\label{Fig4}
\end{figure*}

\begin{figure}
\centering
\includegraphics[width=\columnwidth]{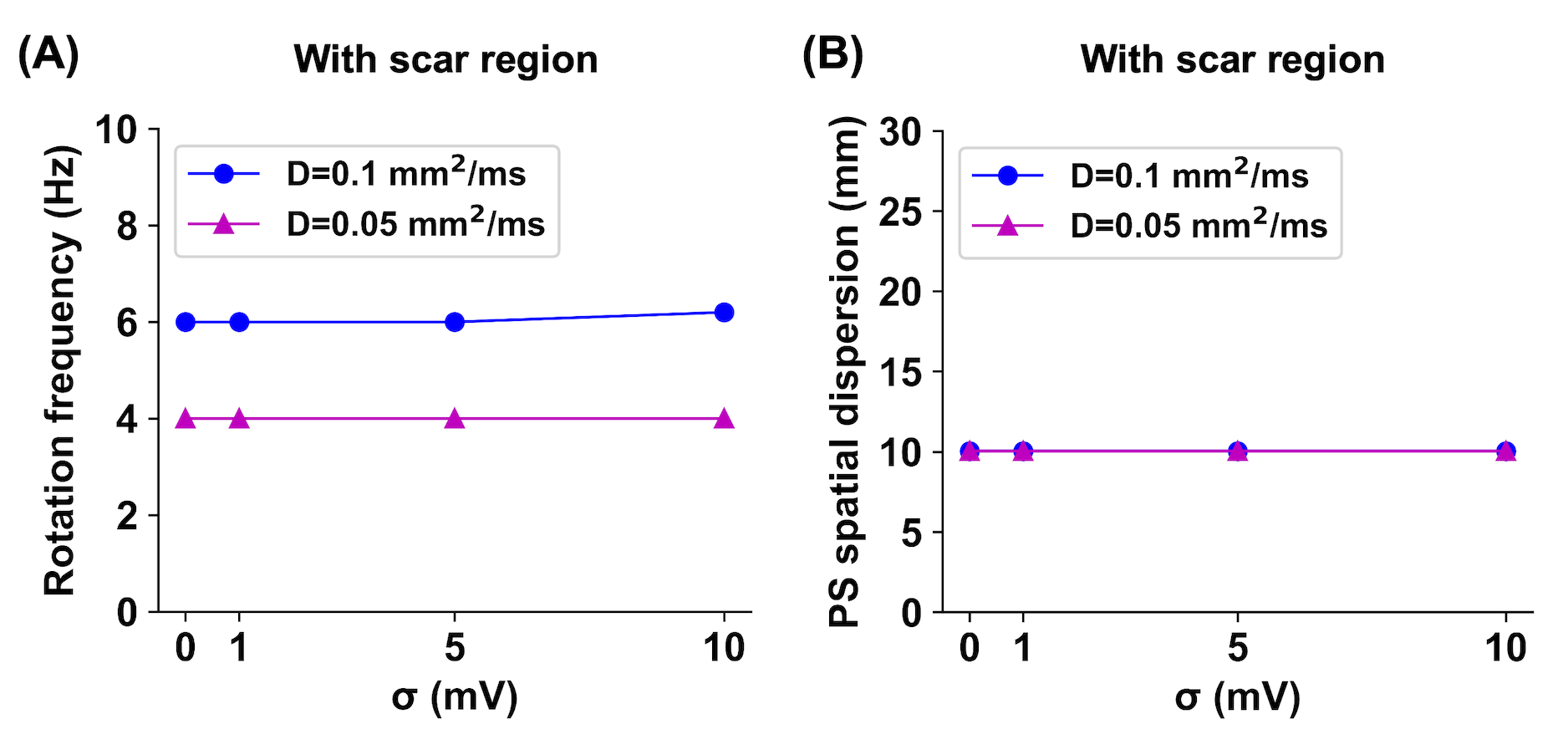}
\caption{Spiral wave rotation frequency (A) and phase singularity (PS) spatial dispersion (B) values for stochastic 2D atrial fibrillation simulations in the presence of scar regions, depending on the diffusion coefficients of $D$=0.1 and 0.05 mm$^2$/ms and the noise intensity levels of $\sigma$=0, 1, 5, and 10 mV.}\label{Fig5}
\end{figure}

\subsection{Effects of scar regions}\label{subsec32}

Next, this study examined the effect of localized scar regions on electrical wave propagation in stochastic AF models. As shown in Figure \ref{Fig4}, electrical waves were periodically propagated around the scar region with a radius of 10 mm. The PS points were identified at the boundary of the scar region. There was no wavefront break-up, and the fibrillation states were sustained for $>$5 s. This stable wave propagation pattern is known as an “anatomical reentry” rather than a “spiral wave,” which is usually defined in the absence of an anatomic obstacle \cite{Allessie:1977}. Noise changed the spiral wave frequencies by only approximately $<$3.3\% (Figure \ref{Fig5}A). As the noise intensity level increased from 0 to 10 mV, the spiral wave rotation frequencies were 6.0, 6.0, 6.0, and 6.2 Hz for the $D$=0.1 mm$^2$/ms cases, and 4.0, 4.0, 4.0, and 4.0 Hz for the $D$=0.05 mm$^2$/ms cases, respectively. In all cases, the PS spatial dispersions were consistently 10.0 mm, which is almost exactly the radius of the scar region (Figure \ref{Fig5}B).

\subsection{Effects of fibrosis regions}\label{subsec33}

\begin{figure*}
\centering
\includegraphics[width=\textwidth]{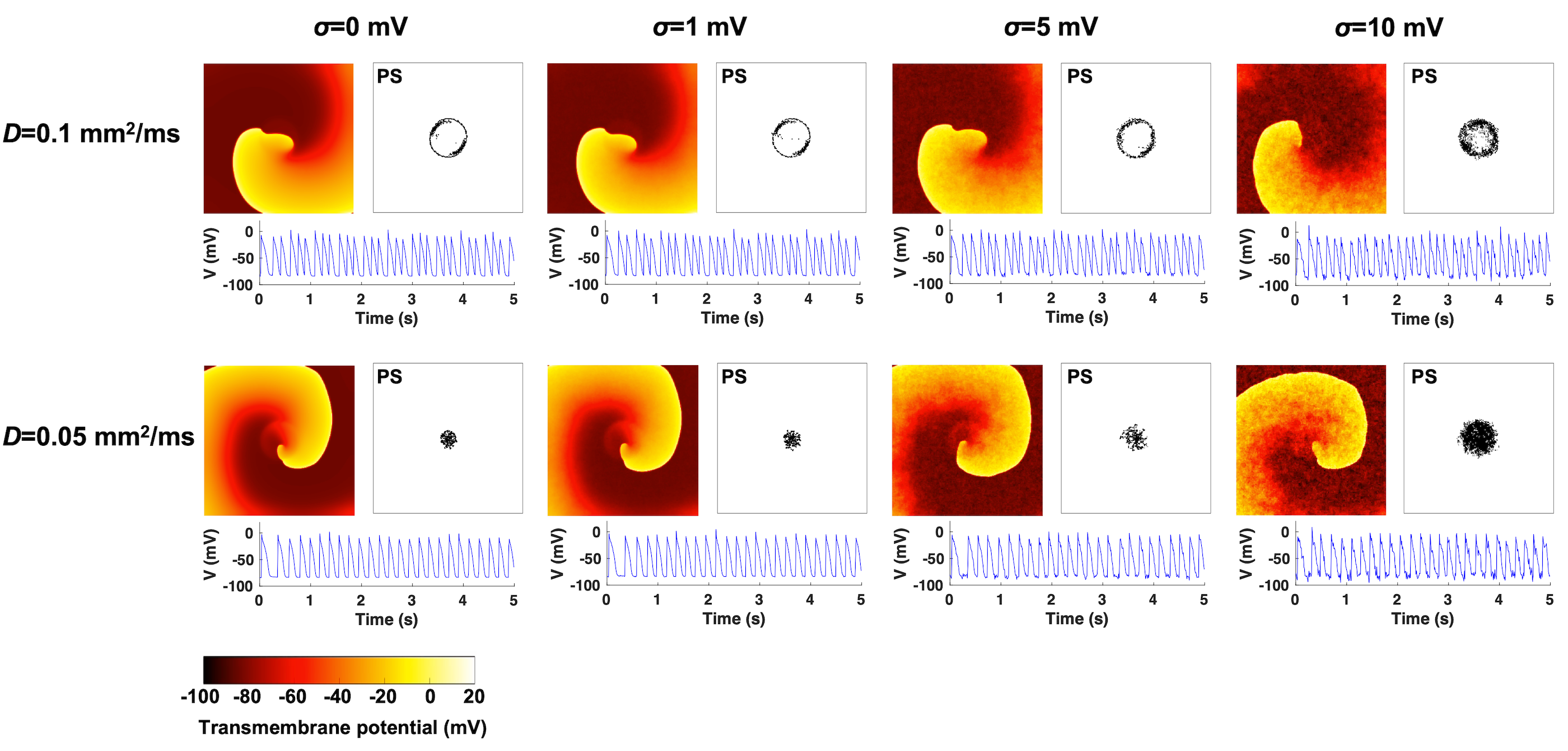}
\caption{Transmembrane potential maps and phase singularity (PS) plots for stochastic 2D atrial fibrillation simulations in the presence of fibrosis regions. The simulations were performed for diffusion coefficients of $D$=0.1 and 0.05 mm$^2$/ms, and noise intensity levels of $\sigma$=0, 1, 5, and 10 mV. The PS points were computed during the whole fibrillation state, and the action potential signals were acquired at the node (50, 50).}\label{Fig6}
\end{figure*}

\begin{figure}
\centering
\includegraphics[width=\columnwidth]{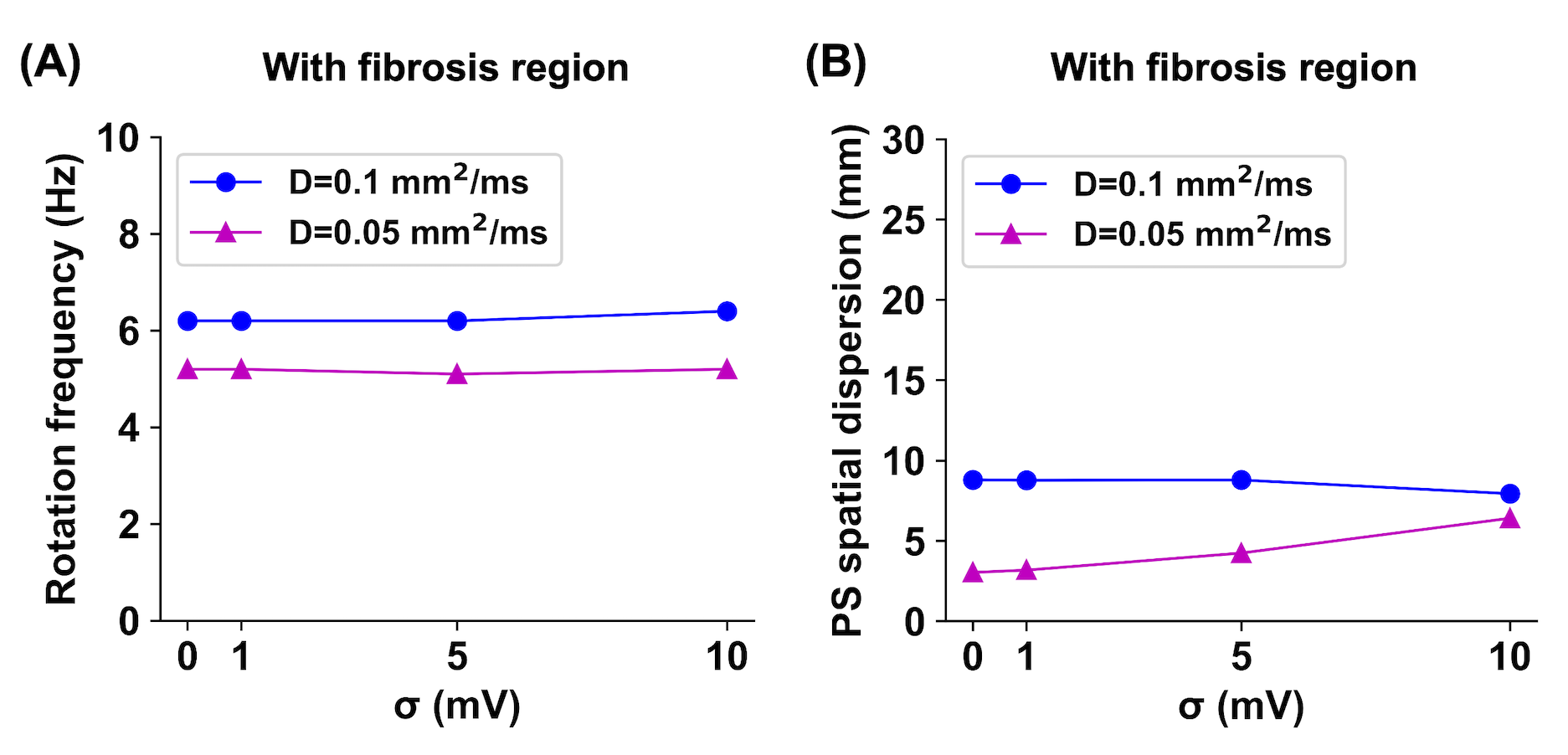}
\caption{Spiral wave rotation frequency (A) and phase singularity (PS) spatial dispersion (B) values for stochastic 2D atrial fibrillation simulations in the presence of fibrosis regions, depending on the diffusion coefficients of $D$=0.1 and 0.05 mm$^2$/ms and the noise intensity levels of $\sigma$=0, 1, 5, and 10 mV.}\label{Fig7}
\end{figure}

Similarly, this study examined how localized fibrosis regions affect the spiral wave dynamics in stochastic AF models. As shown in Figure \ref{Fig6}, when the diffusion coefficient was $D$=0.1 mm$^2$/ms, spiral waves meandered around the fibrosis region with a radius of 10 mm, occasionally invading the fibrosis region when those cells were recovered from refractory periods. When the diffusion coefficient was $D$=0.05 mm$^2$/ms, spiral waves meandered inside the fibrotic region. The PS points were identified at the boundary and inside the fibrotic region. All cases showed sustained fibrillation states for $>$5 s, and there was no wavefront breakup. The noise changed spiral wave frequencies by only approximately $<$3.2\% (Figure \ref{Fig7}A). As the noise intensity level increased from 0 to 10 mV, the spiral wave rotation frequencies were 6.2, 6.2, 6.2, and 6.4 Hz for the $D$=0.1 mm$^2$/ms cases, and 5.2, 5.2, 5.1, and 5.2 Hz for the $D$=0.05 mm$^2$/ms cases, respectively. In all cases, the PS spatial dispersions were consistently below 10.0 mm, implying the spiral wave meandering inside the fibrotic region (Figure \ref{Fig7}B).

\section{Discussion}\label{sec4}

This study numerically simulated stochastic 2D models of human AF and explored the effects of Gaussian white noise on the instability of spiral waves. In homogeneous atrial tissue models, the electrical remodeling condition of the PeAF can generate stable rotor dynamics in the absence of noise. However, Gaussian white noise can lead to spiral-wave meandering and wavefront breakup without significantly altering the spiral wave frequencies (Figures \ref{Fig2} and \ref{Fig3}). In terms of the rotor hypothesis of AF pathophysiology \cite{Narayan:2012,Pandit:2005}, this finding indicates that the cardiac system stochasticity may play an AF-protective role by destabilizing rotors. By contrast, Gaussian white noise is unlikely to affect spiral-wave instability in the presence of localized scar and fibrosis regions, and the PS points are located at the scar or fibrosis areas (Figures \ref{Fig4}–\ref{Fig7}). Thus, scarring or fibrosis may play a pivotal role in stabilizing spiral waves regardless of the Gaussian white noise. The overall results suggest that tissue heterogeneities such as scars and fibrosis are essential for stabilizing the rotors in stochastic 2D mathematical AF models. Further patient-derived stochastic 3D modeling studies are needed to confirm the role of scar/fibrosis in AF pathophysiology.

The pathophysiological importance of fibrosis in AF has been extensively studied. In patient-derived 3D computational models, the PS points are associated with fibrotic regions \cite{Roney:2016,Zahid:2016}; this spatial relationship between fibrosis and the rotor is robust to the model parameter variability \cite{Deng:2017}. In addition, fibroblast–myocyte coupling can affect the spiral wave dynamics and extracellular electrograms \cite{Ashihara:2012,Zlochiver:2008}; however, this coupling effect was not incorporated in this study. The DECAAF clinical study also demonstrated that the degree of atrial tissue fibrosis is associated with the catheter ablation outcomes in AF \cite{Marrouche:2014}. In contrast, a recent non-invasive electrophysiology mapping system found that rotors are not directly associated with fibrosis in patients with AF \cite{Sohns:2017}. This discrepancy between computational and clinical studies may be attributed to the model parameter uncertainty and the absence of stochasticity. Our stochastic AF modeling approach must be further tested to examine the noise-induced instability of spiral waves in patient-derived 3D AF models that reflect patient-specific anatomy and electrophysiology.

Although $D$=0.1 mm$^2$/ms is a widely used diffusion coefficient value in 2D AF models \cite{Pandit:2005,Xie:2002}, I also tested $D$=0.05 mm$^2$/ms to examine the spiral wave dynamics in a severely remodeled condition in PeAF. The results from the two diffusion coefficients were similar, except for the spiral wave frequencies. In homogeneous tissues, the spiral wave frequency is known to be primarily dependent on the inverse of the action potential duration and not on the conduction velocity if the curvature effects are negligible \cite{Qu:2000,Zimik:2020}. In the present study (Figure \ref{Fig3}), the spiral wave frequencies in the homogeneous models were 7.6–8.0 Hz, not being significantly affected by the diffusion coefficient and noise intensity level. However, in the presence of scar or fibrosis, the spiral wave frequencies in the $D$=0.05 mm$^2$/ms cases were consistently lower than those in the $D$=0.1 mm$^2$/ms cases (Figures \ref{Fig5} and \ref{Fig7}). As the PS points were identified at the boundary and inside the scar/fibrosis region, the spiral wave frequency may be mainly dependent on the conduction velocity. Although the spiral wave frequency was not the primary focus of the present study, further studies are needed to systemically examine whether noise alters the spiral wave frequency under various electrophysiological conditions \cite{Mulimani:2022}.

Notwithstanding, this study has several limitations. This study adopted Gaussian white noise, which is neither structurally correlated nor bounded. Because Gaussian noise is inappropriate for many real complex biological systems, the impact of non-Gaussian noise must be investigated \cite{Bobryk:2005,Yao:2017}. The action potential and spiral wave dynamics are also sensitive to the model parameter uncertainty/variability \cite{Mirams:2016,Qu:2000}. The effects of various AF remodeling conditions, tissue anisotropy, and electrophysiological heterogeneity should be systemically investigated further. This study did not consider the noises related to the maximal conductance values and gating variables of ion channels, which are also linked to membrane potential and ion concentrations. Those complex correlated noises across ion channels and membrane potential should be rigorously investigated. This study only tested the noise intensity levels of $\sigma$=0, 1, 5, and 10 mV because of the large computational time. It is worthwhile to determining whether there is a critical $\sigma$ value where transitioning of the instability of spiral waves occurs. In addition, various sizes of atrial tissue and scar/fibrosis regions should be tested because wavelength and tissue size affect the spontaneous termination of cardiac fibrillation \cite{Qu:2006}.

\begin{acknowledgments}
This research received no external funding. This computational study did not produce new animal/clinical data. The author would like to thank the anonymous reviewers for their valuable comments and Editage for English language editing.
\end{acknowledgments}

\section*{Conflict of interest}

The author has no conflicts of interest to declare.

\section*{Authorship contribution}

\textbf{Euijun Song:} Conceptualization, Methodology, Formal analysis, Software, Investigation, Visualization, Writing – original draft.

\nocite{*}
\bibliography{references}

\newpage

\renewcommand{\figurename}{Supplementary Figure}
\setcounter{figure}{0}

\begin{figure*}
\centering
\includegraphics[width=\textwidth]{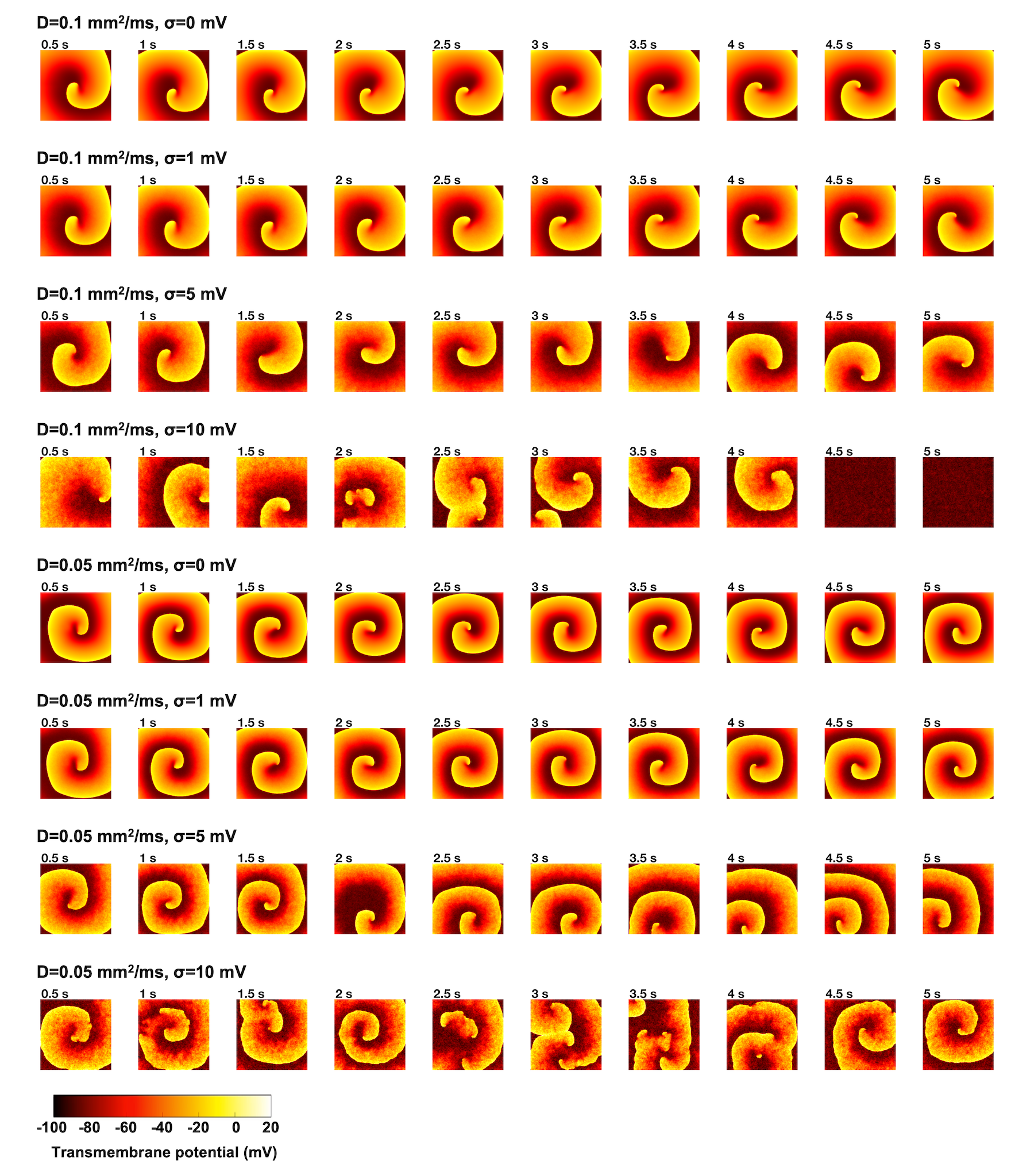}
\caption{Sequences of transmembrane potential maps for stochastic 2D atrial fibrillation simulations on homogeneous tissues.}\label{FigS1}
\end{figure*}

\end{document}